\documentclass[final]{l4dc2020}

\makeatletter
\def\set@curr@file#1{\def\@curr@file{#1}} 
\makeatother
\usepackage[load-configurations=version-1]{siunitx} 

\usepackage{enumerate}
\usepackage{booktabs}
\usepackage[english]{babel}
\usepackage{color}
\usepackage{bm}
\usepackage{times}
\usepackage{caption}
\newcommand{\fb}[1]{#1}

\newcommand{\sss}[1]{_{\scriptscriptstyle #1}}
\newcommand{\uss}[1]{^{\scriptscriptstyle #1}}
\title{LSTM Neural Networks: Input to State Stability and Probabilistic Safety Verification}

\author{\Name{Fabio Bonassi} \Email{fabio.bonassi@polimi.it}\\
\Name{Enrico Terzi} \Email{enrico.terzi@polimi.it} \\
\Name{Marcello Farina} \Email{marcello.farina@polimi.it}\\
\Name{Riccardo Scattolini} \Email{riccardo.scattolini@polimi.it} \\
\addr{Dipartimento di Elettronica, Informazione e Bioingegneria, Politecnico di Milano, via Ponzio 34/5, 20133 Milano, Italy}}

\begin{document}

\maketitle

\begin{abstract}
	The goal of this paper is to analyze Long Short Term Memory (LSTM) neural networks from a dynamical system perspective. The classical recursive equations describing the evolution of LSTM can be recast in state space form, resulting in a time-invariant nonlinear dynamical system.
	A sufficient condition guaranteeing the Input-to-State (ISS) stability property of this class of systems is provided.
	\fb{The ISS property entails the boundedness of the output reachable set of the LSTM. In light of this result, a novel approach for the safety verification of the network, based on the Scenario Approach, is devised.}
	The proposed method is eventually tested on a \emph{pH} neutralization process.
\end{abstract}

\begin{keywords}
	\fb{LSTM}, Input to State Stability, Safety Verification, Scenario Approach.
\end{keywords}

\section{Introduction} \label{sec:intro}
In recent years Neural Networks (NN) have spread as a powerful tool for multiple purposes in several fields of science and engineering \citep{haykin1994neural}, such as system diagnosis \citep{filippetti1995neural} and time series forecasting \citep{azoff1994neural}.
As discussed in \citep{mandic2001recurrent}, among the variants proposed in the literature, Recurrent Neural Networks (RNNs) are of particular interest in' systems identification and control, owing to their ability to describe dynamical systems.


This flexibility, however, comes at the cost of more complex training algorithms \citep{pascanu2013difficulty}, which usually suffer of the vanishing or exploding gradient problems \citep{hochreiter1998vanishing}.
To overcome this issue, novel recurrent architectures have been recently proposed.
Among them, \textit{Echo State Networks} (ESNs) \citep{jaeger2002tutorial, armenio2019model}, and \textit{Long-Short Term Memory} (LSTMs) networks \citep{hochreiter1997long} are considered to be the most promising ones.
More specifically, the latter have gained a great success and are considered to be the state-of-the-art for many tasks entailing long-term sequential learning \citep{greff2016lstm}.

Despite the wide popularity of LSTMs, their use in the context of analysis and control of dynamical systems is still limited, and the associated properties, such as the stability of the equilibria, have not yet been analyzed in depth.
Indeed, only a limited number of contributions is nowadays available concerning stability \citep{stipanovic2018some, deka2019global, barabanov2002stability, miller2018stable}, equilibria computation \citep{amrouche2018long}, and industrial applications \citep{lanzetti2019recurrent}.
In the mentioned works, however, the focus is on autonomous LSTMs, i.e. in absence of exogenous inputs, and the stability analyses conducted -- being based on linearization methods -- are only local.
This essential lack of results needs to be filled in order to provide a solid ground for developing sound control techniques tailored for LSTM.

A further requirement is the ability of LSTMs, and more in general of RNNs, to properly generalize the behavior of the real system out of a limited training set.
It is advisable to exploit the available knowledge of the system to certify that the output produced by the trained neural network is always \fb{coherent with the output of the real system}, at least for a class of inputs of interest.
This is known as \emph{safety verification} of the network, and it is usually done by proving that the output reachable set lays in a defined safe region \citep{maler2008computing}.
The computation of reachable sets for generic discrete-time nonlinear systems has been discussed in \citep{bravo2006robust}.
Techniques to estimate the reachable output set of NNs have been proposed in  \citep{fazlyab2019probabilistic} and \citep{fazlyab2019safety}, but they are limited to feed-forward NNs and cannot be applied to LSTMs.

In light of these motivations, the purpose of this paper is twofold.
First, a sufficient condition guaranteeing the Input-to-State Stability (ISS) property \citep{sontag1995input, jiang2001input} of LSTMs is derived.
\fb{ISS is important here to guarantee consistency when we identify models of plants that display stability-like properties. Also, ISS implies that the state trajectories generated by bounded inputs remain bounded, which is the main assumption that provides a ground to the safety verification approach considered here, based on output reachable sets. The latter sets are estimated, in this paper, using the Scenario Approach introduced by \cite{campi2009scenario}.}
\fb{While a similar idea has been proposed by \cite{hewing2019scenario}, where it is used to estimate the reachable set of a linear system affected by stochastic disturbances, in this paper the Scenario Approach is used in a non-linear deterministic framework, where analytical approaches could not be applied.}
\fb{In a different context, the use of the Scenario Approach for optimization of ESNs has also been proposed by \cite{armenio2019scenario}}.


\medskip
\textbf{Notation}: Given a vector $v$, we denote with $|v|\sss{2}$ the 2-norm of $v$ and with $|v|\sss{1}$ its $1$-norm.
Moreover, we denote with $v^T$ its transpose, with $v_i$ its $i$-th component, and with ${\rm diag}(v)$ the diagonal matrix with vector $v$ on the main diagonal.
With reference to a sequence of vectors $\mathbf{v}=v(0), v(1), ...$, we define its infinity norm as $\|\mathbf{v}\|\sss{\infty}=\sup_{k \in \mathbb{N}}|v(k)|\sss{2}$.
Given a real matrix $A$ we denote with $|A|_p$ its induced norm, $p=1,2,\infty$.
At any discrete time-step $k$, we indicate by $x\uss{+}$ the value of $x$ at the next step, i.e. $x\uss{+} = x(k+1)$.
The Hadamard product between two vectors $v$ and $w$ is indicated by $v \circ w$.
For compactness, the sigmoidal and hyperbolic tangent activation functions are denoted as $\sigma_g(x)=\frac{1}{1+e^{-x}}$ and $\sigma_c(x)={\rm tanh}(x)$.
These functions are bounded, i.e.
\begin{subequations} \label{eq:actbounds}
	\begin{align}
		\sigma_g(t) \in (0,1) \quad & \forall t  \in \mathbb{R}, \label{eq:bound_sigmag} \\
		\sigma_c(t) \in (-1,1) \quad & \forall t \in \mathbb{R}, \label{eq:bound_sigmac}
	\end{align}
\end{subequations}
and Lipschitz continuous \citep{sohrab2003basic}, with constants $L_g=0.25$ and $L_c=1$, respectively.
%

\section{LSTM networks}\label{sec:model}
\subsection{LSTM state space form}
Let us consider a single-layer LSTM network described by the following equations, \fb{obtained enforcing the original equations from} \citep{gers2001lstm} \fb{to be strictly proper}: 
\begin{subequations} \label{eq:statespace}
	\begin{align}
		x\uss{+}	 & = \, 	\sigma_g(W_f u + U_f \xi + b_f) \circ x+ \sigma_g(W_i u + U_i \xi + b_i) \circ \sigma_c(W_c u + U_c \xi + b_c)\label{eq:state_eq_orig}\\
		\xi\uss{+}	 &= \, \sigma_g(W_o u + U_o \xi + b_o) \circ \sigma_c(x\uss{+}) 		\label{eq:output_orig},\\
		y &= \, C \xi + b_y , \label{eq:output_orig1}
	\end{align}
\end{subequations}
where $u \in \mathbb{R}^{n_u}$ is the input, $x \in \mathbb{R}^{n_x}$ is the so-called hidden state, $\xi \in \mathbb{R}^{n_x}$ is the so-called cell state, and $y\in \mathbb{R}^{n_y}$ is the system output.
The tuning parameters, suitably selected during network's training procedure, are the weight matrices $W_f,W_i,W_c,W_o \in \mathbb{R}^{n_x \times n_u}$, $C\in \mathbb{R}^{n_u,2n_x}$, $U_f,U_i,U_c,U_o \in \mathbb{R}^{n_x \times n_x}$, and the bias vectors $b_f,b_i,b_c,b_o,b_y$ of proper dimensions.
In this paper, without loss of generality, inputs are assumed to be bounded by
\begin{subequations} \label{eq:bounds}
	\vspace{-1mm}
	\begin{equation} \label{satu}
		u \in \mathcal{U} = [-1,1]^{n_{u}}.\vspace{-1mm}
	\end{equation}
Indeed, the inputs are usually subject to saturation, and can therefore be normalized according to standard un-biasing and normalization procedures.
In light of \eqref{eq:actbounds}, from \eqref{eq:output_orig} it is possible to notice that the vector of cell states is also bounded by\vspace{-1mm}
\begin{equation}
	\xi\in \Xi = (-1,1)^{n_x}.\vspace{-1mm}
\end{equation}
\end{subequations}
At this stage, let us consider the bias vector $b_c$ as a constant input to system \eqref{eq:statespace}. \fb{This is merely required for the sake of consistency with the existing theory, and indeed the bias $b_c$ is commonly identified during the training of the network}.
Letting $\chi = \begin{bmatrix}x^T  & \xi^T\end{bmatrix}^T$, it is possible to re-write \eqref{eq:statespace} in the standard form of a discrete-time, invariant, and nonlinear dynamical system:
\vspace{-1mm}
\begin{equation}\label{eq:LSTMmodel}
	\begin{aligned}
	\chi\uss{+} &= f_{\scriptscriptstyle LSTM}(\chi,u,b_c), \\
	y &= g_{\scriptscriptstyle LSTM}(\chi). \\
	\end{aligned}\vspace{-1mm}
\end{equation}

\subsection{Stability analysis}
The goal of this section is to investigate the stability properties of the LSTM network \eqref{eq:LSTMmodel}.
More specifically, we provide a sufficient condition to guarantee the ISS property of the network.
Let us firstly recall some notions from \citep{bayer2013discrete}, required for the statement of the theoretical results of this paper.
For the definitions of $\mathcal{K}, \, \mathcal{K}_{\infty}$ and $\mathcal{KL}$ functions, the reader is addressed to \citep{jiang2001input}.
Considering the sequence $\mathbf{u}=u(0),u(1),\ldots$ and an initial state $\chi_0$, we denote by $\chi\left(k, \chi_o, \mathbf{u}, b_c \right)$ the state, at time $k$, of system \eqref{eq:LSTMmodel}, fed by the input sequence $\mathbf{u}$.
The following definitions can hence be stated.
%
\begin{definition}[$ISS$ \citep{jiang2001input}]
	\label{deltaISSdef}
	The system \eqref{eq:LSTMmodel} is \textit{Input-to-state stable} with respect to inputs $u\in\mathcal{U}$ and $b_c$ if there exist functions $\beta\in\mathcal{KL}$, and $\gamma_u,\gamma_b\in \mathcal{K}_{\infty}$ such that, for any $k\in\mathbb{Z}_{\geq0}$, any initial condition $\chi_{0}$, any value of $b_c$, and any input sequence $\mathbf{u}\in\mathcal{U}$, it holds that:
	\begin{equation}  \label{eq:ISS}
		|\chi(k,\chi_{0},\mathbf{u}, b_c)|\sss{2}\leq\beta(|\chi_{0}|\sss{2},k)+\gamma_u(\|\mathbf{u}\|\sss{\infty})+\gamma_b(|b_c|\sss{2})
	\end{equation}
\end{definition}
\begin{definition}[ISS Lyapunov function \citep{jiang2001input}]
	A continuous function $V:\mathbb{R}^{n}\to\mathbb{R}_{+}$ is called an    ISS-Lyapunov function for \eqref{eq:LSTMmodel} if there exist functions $\psi_{1},\psi_{2},\psi\in\mathcal{K}_{\infty}$ and $\sigma_u,\sigma_b\in\mathcal{K}$ such that, for all $\chi\in\mathbb{R}^{2n_x}$, for all $b_c\in\mathbb{R}^{n_x}$, and $u\in\mathbb{R}^{n_u}$, it holds that:
	\begin{subequations}
		\begin{align}
		\psi_{1}(|\chi|\sss{2})\,\,\leq \,\, &V(\chi) \leq \,\,\psi_{2}(|\chi|\sss{2})\label{firstCond1}\\
		V(f_{\scriptscriptstyle LSTM}(\chi,u,b_c)) \,- \,& V(\chi)\leq-\psi(|\chi|\sss{2})+\sigma_u(|u|\sss{2})+\sigma_b(|b_c|\sss{2})\label{secondCond1}
		\end{align}
	\end{subequations}
\end{definition}

\begin{theorem}[\cite{jiang2001input}]
	\label{Teo:ISSLyap}
	If system \eqref{eq:LSTMmodel} admits a time invariant $ISS$ Lyapunov function such that \eqref{firstCond1} and \eqref{secondCond1} hold, then it is $ISS$ in the sense specified by Definition~\ref{deltaISSdef}.
\end{theorem}
Based on the previous results, Theorem \ref{thm:ISS} is formulated. The proof is reported in Appendix \ref{proof:ISS}.
\begin{theorem}\label{thm:ISS}
	Given the LSTM network \eqref{eq:statespace}, if\vspace{-1mm}
	\begin{equation}
	\begin{array}{ll}
	(1+\sigma_g(|\begin{bmatrix}W_o&U_o&b_o\end{bmatrix}|\sss{\infty})) \, \sigma_g(|\begin{bmatrix}W_f&U_f&b_f\end{bmatrix}|\sss{\infty})&<1, \\
	(1+\sigma_g(|\begin{bmatrix}W_o&U_o&b_o\end{bmatrix}|\sss{\infty})) \, \sigma_g(|\begin{bmatrix}W_i\,&U_i\,&b_i\,\end{bmatrix}|\sss{\infty})\,|U_c|\sss{1}&<1,
	\end{array}\vspace{-1mm}
	\label{cond_teo1}
	\end{equation}
	then \eqref{eq:statespace} is Input-to-State stable with respect to $u\in\mathcal{U}$ and to $b_c$.
\end{theorem}
\begin{remark}
	Condition \eqref{cond_teo1} involves the LSTM's weight matrices and bias vectors.
	Therefore, it can be employed to check a-posteriori the ISS of a trained LSTM network, or it can be used as a constraint during the training procedure of the network to enforce the ISS property.
\end{remark}
\begin{remark}
	\fb{If multi-layer LSTMs are considered, it should be noted that the cell state $\xi$ of the $i$-th layer becomes the input of the next layer. Since $\xi$ satisfies the same properties input $u$, namely \eqref{satu}, it follows that a sufficient condition for the ISS of the  network is that each layer satisfies \eqref{cond_teo1}.}
\end{remark}

\section{Safety verification} \label{sec:verification}
As discussed, training the LSTM network with constraints \eqref{cond_teo1} explicitly enforced, allows one to obtain the  ISS property.
\fb{This property is extremely relevant in the context of systems control, as it guarantees the boundedness network's reachable set, and that the effect of initialization on network's output asymptotically vanishes.}
Nonetheless, the reliability of the network could be harmed by the limited size of the training set usually adopted, either due to data unavailability or to computational complexity reasons.
Overfitting phenomena and limitations of the available numerical training procedures could, as well, undermine LSTMs' generalization capabilities.
This entails the necessity to perform a safety verification of the network, in order to certify -- at least for a given class of input signals -- that it does not exhibit meaningless or non-physical outputs.
This task is usually accomplished computing the output reachable set, and assessing that it entirely lays within a known safe set.
While for feed-forward NNs some methods for the estimation of the output reachable set have been proposed, see \citep{fazlyab2019safety}, for RNNs this still represents an open problem.

First remark that, in view of Definition \ref{deltaISSdef}, the ISS property allows to retrieve an explicit bound on the reachable set of the network.
In particular, for any initial state $\chi_0$, and any bounded input sequence $\nu(0), \nu(1), \ldots$, the future state trajectories are guaranteed to be bounded by \eqref{eq:ISS}, where the functions $\beta(\cdot, \cdot)$, $\gamma_u(\cdot)$, and $\gamma_b(\cdot)$ can be computed from the ISS-Lyapunov function, as described in \citep{jiang2001input}.
The output reachable set can then be computed applying the linear output transformation $g_{\scriptscriptstyle LSTM}$.
Nonetheless, such bound may suffer from an excessive conservativeness.
The tightening of  ISS-based reachable set bound will be object of future research work.

\medskip
In place of the ISS-based analytic (but possibly conservative) computation  of the output reachable set, a probabilistic method is here proposed, based on the Scenario Approach developed in \citep{campi2009scenario}.
Let the initial state of the LSTM network $\chi_0$ be a random variable extracted from a set $\mathcal{X}_0$, characterized by some probability measure $\mathbb{P}_{\chi}$.
Consider a time horizon $\tau$, and a class $\bm{\mathcal{U}}_\tau$ of input signals $u(0), \ldots, u(\tau)$, such that $u(k) \in \mathcal{U}$ for all $k=0, \ldots, \tau$.
Assume that $\bm{\mathcal{U}}_\tau$ is characterized by some -- possibly unknown -- probability measure $\mathbb{P}_u$. 
\fb{To ensure the validity of the results, the class of inputs $\bm{\mathcal{U}}_\tau$ shall be compliant with the policy by which the system is operated.}

The goal is to find the smallest ball $\mathcal{Y}$ containing the output reachable set, i.e. the set where the output trajectories produced by any input sequences extracted from  $\bm{\mathcal{U}}_\tau$ and starting from any initial state drawn from $\mathcal{X}_0$ lie.
The radius $\rho_y^*$ of set $\mathcal{Y}$ is defined as the solution of

\begin{subequations} \label{eq:verification:inf}
	\begin{align}
		\rho_y^*=\min_{\rho_y} \quad & \rho_y, \\
		s.t. \quad &  \| \mathbf{y}(\chi_{0}, \mathbf{u}) \|\sss{\infty} \leq \rho_y  \qquad \forall{\chi_0}\in \mathcal{X}_0, \, \forall \mathbf{u} \in \bm{\mathcal{U}}\tau, \label{eq:verification:inf:constr}
	\end{align}
\end{subequations}
where $\mathbf{y}(\chi_{0}, \mathbf{u})$ is the output sequence obtained feeding the trained LSTM \eqref{eq:LSTMmodel} with the input sequence $\mathbf{u}$ and initial state $\chi_{0}$. 
Notably, this problem cannot be solved directly, due to infinite cardinality of constraint  \eqref{eq:verification:inf:constr}.
Nonetheless, owing to the convexity of \eqref{eq:verification:inf} with respect to the optimization variable $\rho_y$, the Scenario Approach proposed by \cite{campi2009scenario} can be exploited to recast the optimization problem as a finite-dimensional linear program, that allows to compute $\rho_y^*(\varepsilon,\beta)$ such that
\begin{equation}
	\mathbb{P}_{\chi, u} \left\{  \| \mathbf{y}(\chi_{0}, \mathbf{u}) \|\sss{\infty} > \rho_y^*(\varepsilon,\beta) \right\} \leq \varepsilon
\end{equation}
with confidence $1-\beta$. In this way, the deterministic problem \eqref{eq:verification:inf:constr} is relaxed into a chance-constrained one.
To do so, it is necessary to generate a number $N$ of scenarios, each corresponding to a sample of the uncertain variables $\chi_0$ and $\mathbf{u}$, drawn from the respective sets according to the associated probability density functions. We define with $\chi_0^{(i)},\mathbf{u}^{(i)}$ the $i$-th scenario, where $i=1,\dots,N$. Theorem 1 of \citep{campi2009scenario} allows to compute $\rho_y^*(\varepsilon,\beta)$ as
\begin{equation} \label{eq:verification:scenario}
\begin{aligned}
	\rho_y^*(\varepsilon,\beta)=\min_{\rho_y} \quad & \rho_y, \\
	s.t. \quad &  \| \mathbf{y}(\chi_{0}^{(i)}, \mathbf{u}^{(i)}) \|\sss{\infty} \leq \rho_y  \,\,\, \text{ for all } i=1, ..., N,
\end{aligned}
\end{equation}
provided that a sufficient number $N$ of scenarios is adopted.
In particular, being $d=1$ the number of optimization variables, it is sufficient that $N$ fulfills the following inequality \citep{campi2009scenario}.
\begin{equation} \label{eq:verification:Nscen}
	N \geq \frac{2}{\varepsilon} \left( \ln\frac{1}{\beta} + d \right)
\end{equation}

\vspace{-3mm}
It is worth remarking that \eqref{eq:verification:scenario} is linear and convex, since $ \mathbf{y}(\chi_{0}^{(i)}, \mathbf{u}^{(i)})$ does not depend on any optimization variable. To conclude, defining $\mathcal{Y}(\varepsilon,\beta) = \{ y \in \mathcal{R}^{n_y}: |y|\sss{2} \leq \rho_y^*(\varepsilon,\beta)\}$, if $\mathcal{Y}$ is contained in the safe output set, then one can assess - with the mentioned probability levels - the safety of the identified LSTM network, \fb{as long as the inputs belong to the considered class.}

\section{Simulation example} \label{sec:benchmark}
\begin{figure}
	\centering
	\includegraphics[clip, trim=0cm 0.25cm 0cm 0.2cm, width=0.35 \linewidth]{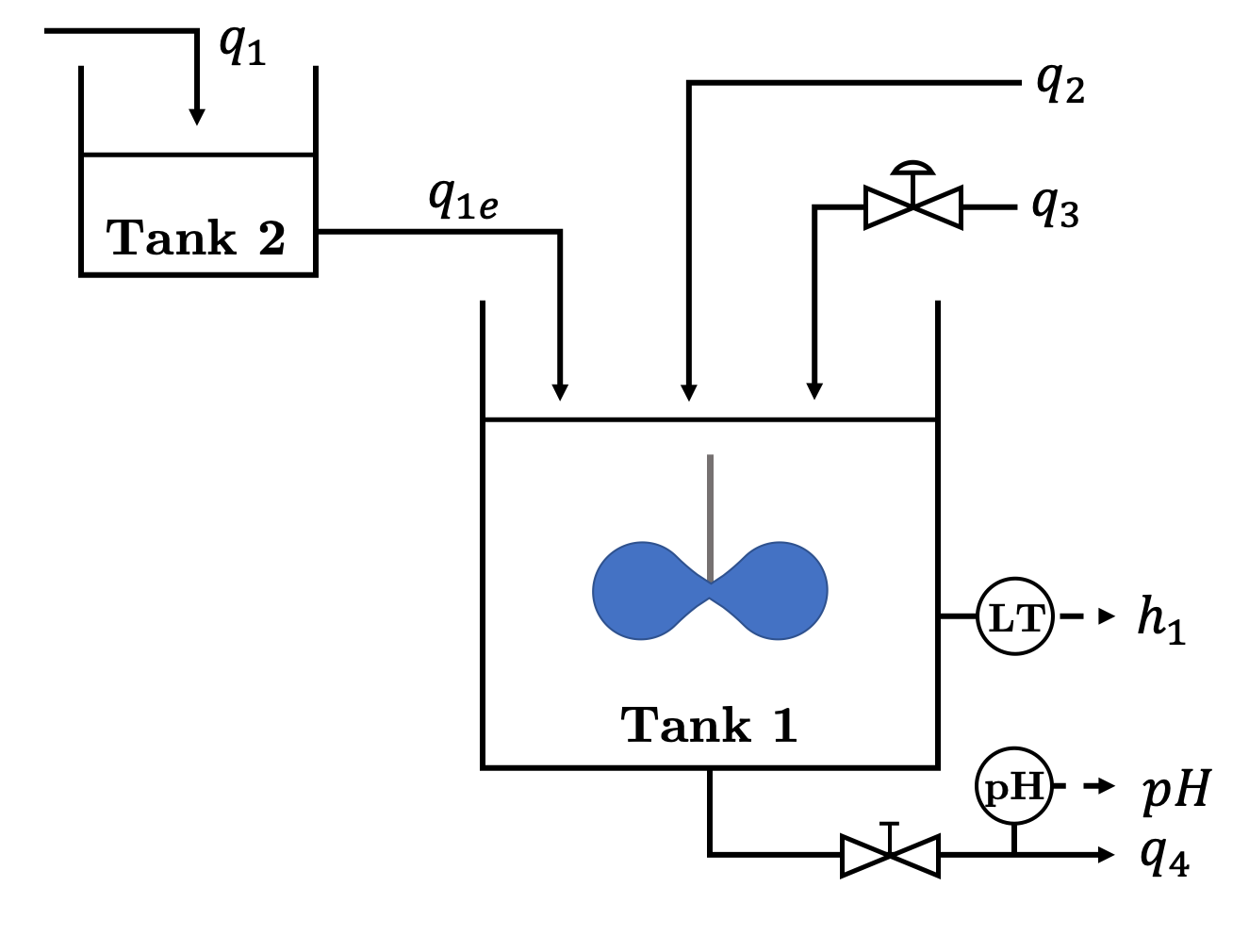}
	\vspace{-0.25cm}
	\captionof{figure}{Schematic layout of the plant}
	\label{fig:PhScheme}
\end{figure}
The considered benchmark example is the \emph{pH} neutralization process described in \citep{hall1989modelling} and  used in \citep{armenio2019model} for the analysis of ESNs. The plant, depicted in Figure \ref{fig:PhScheme}, is composed of two tanks.
Tank 2 is fed by an acid stream $q_1$ and, as output, has the flowrate $q_{1e}$. 
Being the hydraulic dynamics faster than the others involved, we assume $q_{1}=q_{1e}$. 
The reactor Tank 1 is fed by three flowrates, namely $q_1$, a buffer flowrate $q_2$ and an alkaline flowrate $q_3$.
$q_1$ and $q_2$ are not manipulated variables, and represent disturbances, while a controlled valve modulates $q_3$. 
The output flowrate of the tank is $q_4$, where the \emph{pH} is measured.

The simplified third-order model, see \citep{armenio2019model} for equations, has one input ($q_3$), and one output (the $pH$ value).
The simulator of this plant has been implemented in MATLAB, and it has been fed with a Multilevel Pseudo-Random Signal (MPRS), so as to properly excite the system, adopting a sampling time $T_s=10 \text{s}$.
White noise has been added, both to input and to measured output data, to mitigate overfitting phenomena.
The training and validation sets consist of $4400$ and $2250$ $(u, y)$ samples, respectively. 
Samples have been normalized using the mean and the maximum deviation across the entire dataset.

\begin{figure}
	\centering
	\includegraphics[clip, trim=0cm 0.085cm 0 0.35cm, width=0.35 \linewidth]{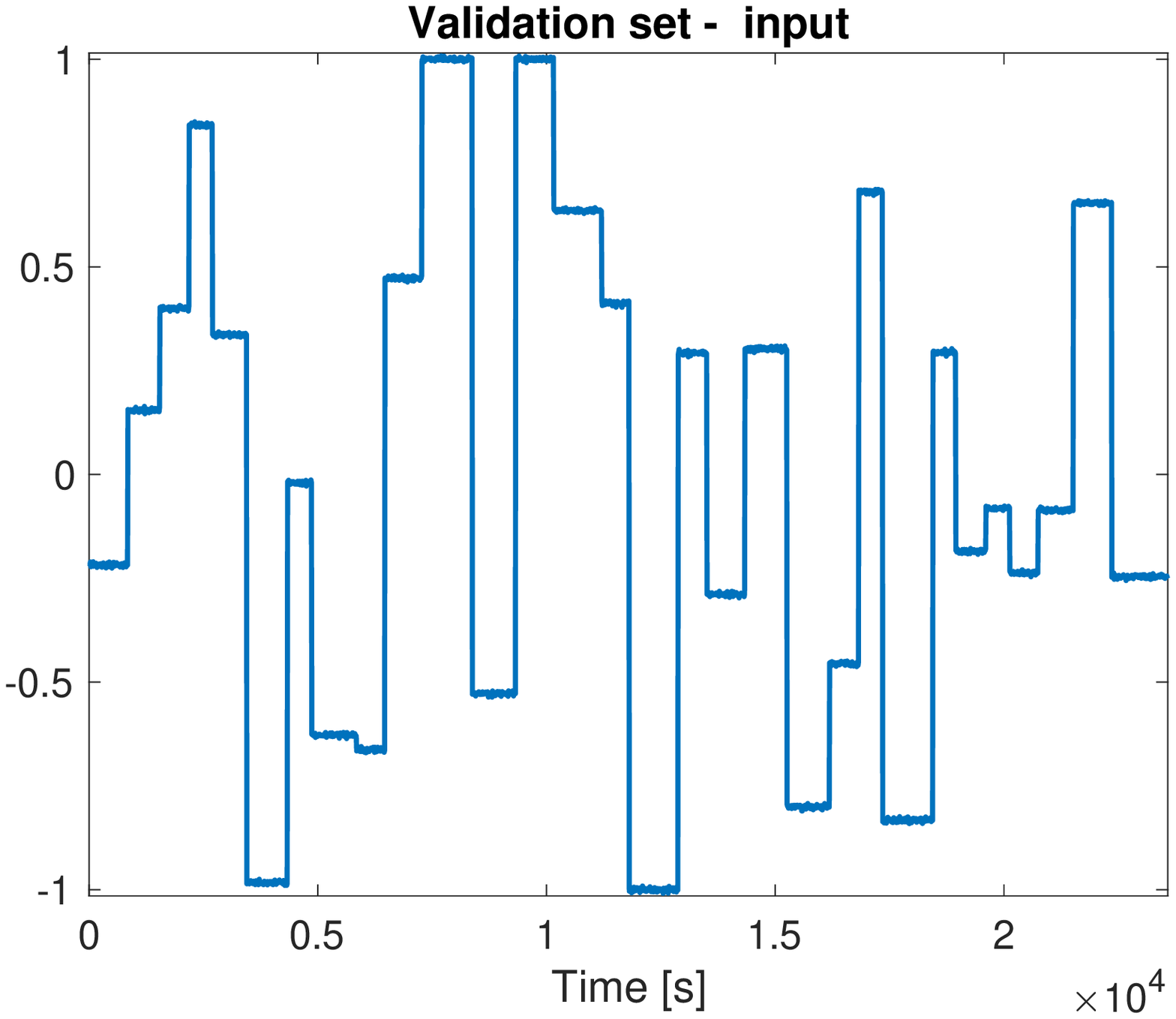}	 
	\includegraphics[clip, trim=0cm 0.085cm 0 0.35cm, width=0.35 \linewidth]{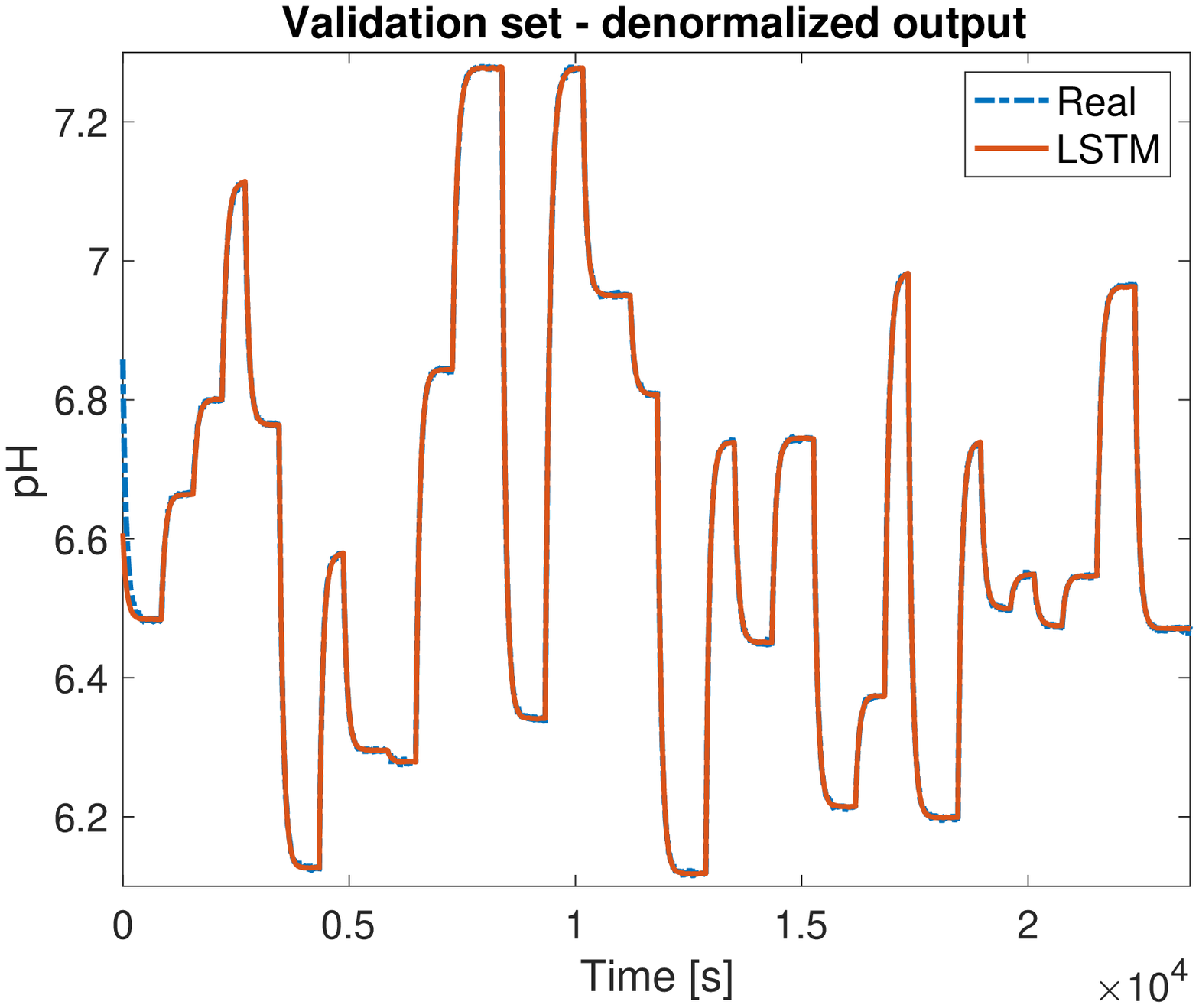}	
	\caption{Performances of the LSTM on the validation dataset.}
	\label{fig:validation}
\end{figure}

A single-layer LSTM neural network \eqref{eq:statespace} with $n_x = 5$ cells has then been trained using the MATLAB function \emph{fmincon} \fb{to minimize the mean-square prediction error.}
In order to ensure network's ISS, conditions \eqref{cond_teo1} have been enforced as explicit constraints during the training procedure.
The modeling performances over the validation dataset are reported in Figure \ref{fig:validation}, where a simulation of the trained network, starting from a random initial state and forced by the input $u$, is shown.
Notice that, in Figure \ref{fig:validation}, the denormalized output is depicted for the sake of clarity.

A quantitative performance index is the FIT $[\%]$ value, computed as
\begin{equation}
\text{FIT} = 100\left(1- \frac{|\mathbf{y}\sss{val} - \mathbf{y}\sss{LSTM}|\sss{2}}{|\mathbf{y}\sss{val}|\sss{2}} \right),
\label{eq:FIT}
\end{equation}
where $\mathbf{y}\sss{val}$ is the real system output trajectory and $\mathbf{y}\sss{LSTM}$ is the output of the trained LSTM network, fed by the same input. The trained network scores, after an initial transient due to the wrong initial state, $\text{FIT} = 98\%$ , confirming remarkable modeling properties.

The probabilistic safety verification proposed in Section \ref{sec:verification} has then been tested on the LSTM network.
In particular, the input class $\bm{\mathcal{U}}_\tau$ selected for this test is the set of MPRS, with amplitude $\rho_u$ and with a total duration of $\tau = 2000$ time-steps, i.e. $20 000$ seconds.
Since $n_u=1$, any sequence $\mathbf{u}^{(i)}$ sampled from $\bm{\mathcal{U}}_\tau$ is composed by steps with an amplitude uniformly extracted from $[ - \rho_u, \rho_u ]$, and with a random duration uniformly drawn in the range $[ 300, 2000 ]$ seconds.
In Figure \ref{fig:verification}\hyperref[fig:verification]{\textcolor{blue}{(a)}} some examples of input sequences thus generated have been depicted, for $\rho_u=0.7$.

The initial state $\chi_0$ has been extracted from a uniform distribution over $ \mathcal{X}_0 = [-0.1, 0.1]^{2 n_x}$. Notice that, owing to LSTM's ISS, the effect of the initial state asymptotically vanishes.

In light of condition \eqref{eq:verification:Nscen}, to have a violation probability $\varepsilon = 10^{-2}$ with confidence $\beta = 10^{-6}$, $N=2964$ scenarios must be generated.
Problem \eqref{eq:verification:scenario} is hence evaluated for different values of $\rho_u$, namely for $\rho_u = \{ 0.1, 0.2, ..., 1\}$, both for the LSTM network and the real system.
Remarkably, the results reported in Figure \ref{fig:verification}\hyperref[fig:verification]{\textcolor{blue}{(b)}} witness that the LSTM is guaranteed to operate almost in the same region of the real system, i.e. a ball with radius $\rho_y \approx 1$. 

\begin{figure}
\floatconts
  {fig:verification}
  {\vspace{-0.75cm} \caption{(a) Some examples of input sequence scenarios for $\rho_u = 0.7$; (b) Bounds $\rho_y^*$ on the LSTM normalized output reachable set for different input bounds $\rho_u$, compared to those of the  real system.}}
  {
    \subfigure[]{
      \includegraphics[clip, trim=0.15cm 0cm 0.15cm 0cm, width=0.375  \linewidth]{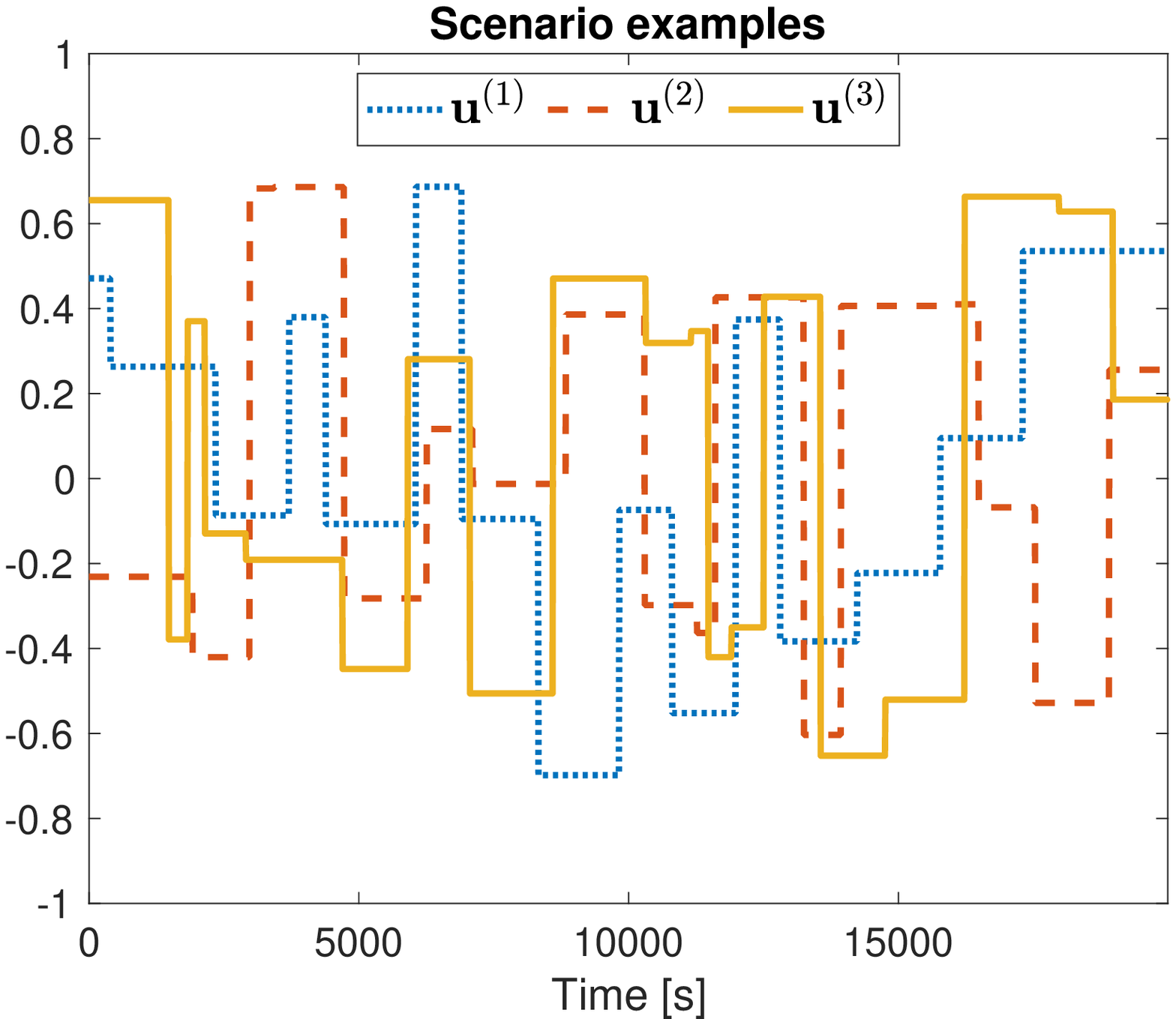}}    \quad
    \subfigure[]{\includegraphics[clip, trim=0.25cm 0.35cm 0.25cm 0cm, width=0.375 \linewidth]{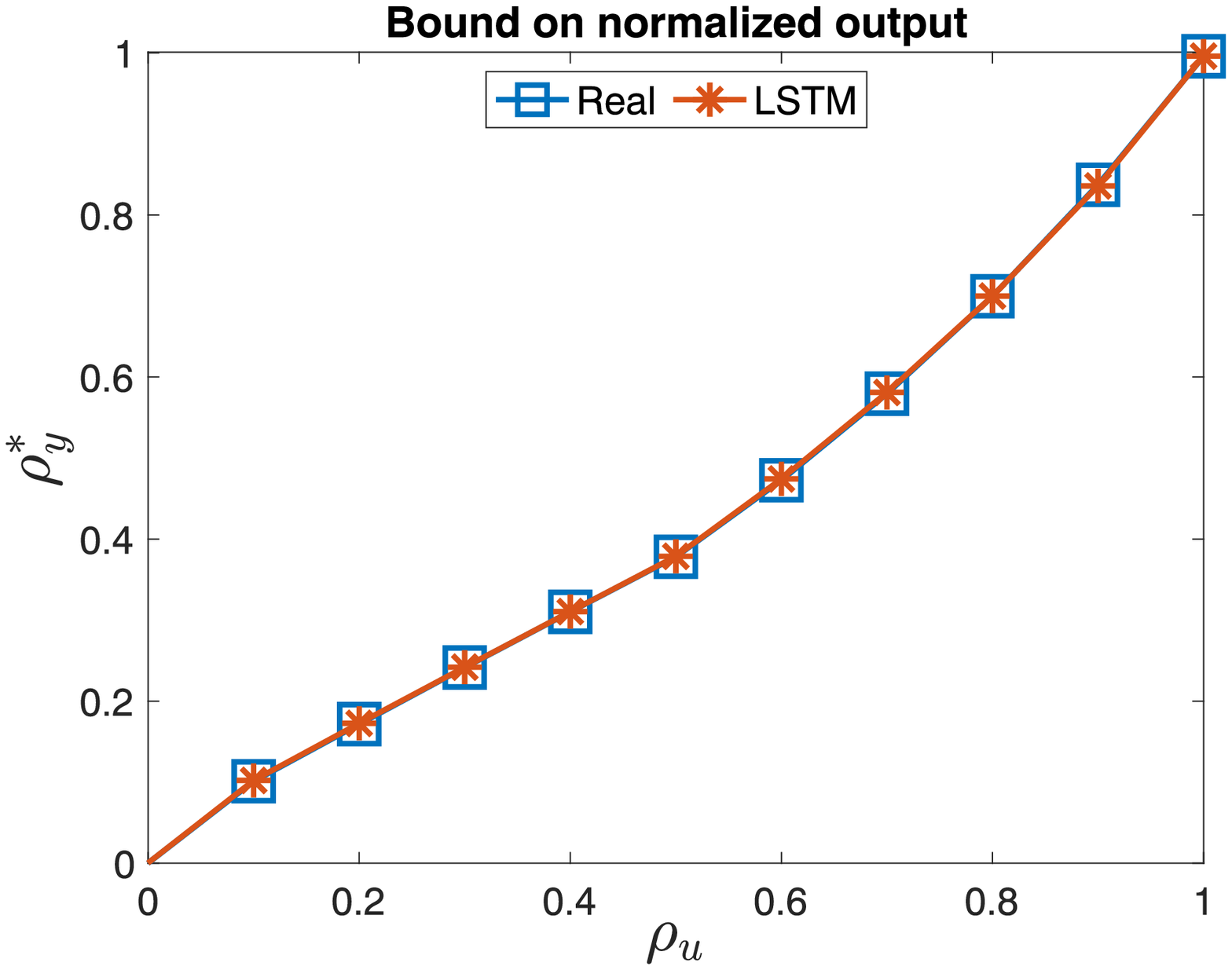}}
  }
\end{figure}

\section{Conclusions}
In this work, the stability properties of LSTM neural networks have been investigated.
In particular, a condition on LSTM's parameters to guarantee the Input-to-State Stability property of the network has been derived, which can be possibly enforced during the training procedure.
A probabilistic method for the safety verification of the trained network has then been formulated, which allows -- for any class of input signals -- to retrieve a tight bound on network's output reachable set, up to a desired confidence level.
The proposed algorithm has been eventually tested on a well-known benchmark system, the \emph{pH} neutralization process, showing satisfactory performances.

\appendix
\section{Proof of Theorem \ref{thm:ISS}} \label{proof:ISS}
Let us define the following variables, for the sake of conciseness:
\begin{equation} \label{eq:aux_def}
\begin{aligned}
	f(u,\xi) &= W_f u + U_f \xi + b_f,  \qquad
	& i(u,\xi) &= W_i u + U_i \xi + b_i, \\
	c(u,\xi) &= W_c u + U_c \xi + b_c, \qquad
	& o(u,\xi) &= W_o u + U_o \xi + b_o.
\end{aligned}
\end{equation}
By means of these definitions, the state equations \eqref{eq:state_eq_orig} and \eqref{eq:output_orig} can be written as
\begin{subequations}
	\begin{align}
		x\uss{+}	 = & 	\sigma_g(f(u,\xi)) \circ x+ \sigma_g(i(u,\xi)) \circ \sigma_c(c(u,\xi)), \label{eq:state}\\
		\xi\uss{+}	 = & \sigma_g(o(u,\xi)) \circ \sigma_c(x\uss{+}). \label{eq:output}
	\end{align}\label{eq:statespace_new}
\end{subequations}
$\!\!\!$ Letting $\alpha\sss{f}(u,\xi) \! = \! {\rm diag} \left ( \sigma_g(f(u,\xi))	\right)$, $\alpha\sss{o}(u,\xi) \! = \! {\rm diag} \left ( \sigma_g(o(u,\xi))	\right)$, and $\alpha\sss{i}(u,\xi) \! = \! {\rm diag} \left ( \sigma_g(i(u,\xi)) \right)$, it is possible to re-write \eqref{eq:statespace_new} as
\begin{align}\label{eq:LPV}
	\begin{bmatrix}
	x\uss{+}\\
	\xi\uss{+}
	\end{bmatrix}= \begin{bmatrix} \alpha\sss{f}(u,\xi) \, x + \alpha\sss{i}(u,\xi) \, \sigma_c(c(u,\xi))\\
	\alpha\sss{o}(u,\xi) \, \sigma_c\big( \, \alpha\sss{f}(u,\xi) \, x + \alpha\sss{i}(u,\xi) \, \sigma_c(c(u,\xi)) \,\big)
	\end{bmatrix}.
\end{align}

In light of Theorem \ref{Teo:ISSLyap}, consider the following candidate ISS-Lyapunov function:
\begin{equation} \label{eq:Lyap_function}
V(\chi)=|\chi|\sss{1}=|x|\sss{1}+|\xi|\sss{1}.
\end{equation}
By definition $V(\chi) \geq 0, \forall \chi\neq 0$. Condition \eqref{firstCond1} follows trivially, since for a generic vector $v \in \mathbb{R}^{n_v}$ it holds that $|v|\sss{2} \leq |v|\sss{1} \leq \sqrt{n_v} \, |v|\sss{2}$. Therefore, condition \eqref{firstCond1} is fulfilled taking
\begin{align*}
	\psi_1(|\chi|\sss{2})&=|\chi|\sss{2}, \\
	\psi_2(|\chi|\sss{2})&=\sqrt{2n_x}|\chi|\sss{2}.
\end{align*}
We now compute $V(\chi\uss{+}) - V(\chi)$, showing that there exist functions $\psi(|\chi|\sss{2})$, $\sigma_u(|u|\sss{2})$ and $\sigma_b(|b_c|\sss{2})$ fulfilling condition \eqref{secondCond1}.
Using \eqref{eq:LPV} one obtains
\begin{equation*}
	\begin{aligned}
	V(\chi\uss{+}) - V(\chi)  =\, & \lvert \alpha\sss{f}(u,\xi \,) x + \alpha\sss{i}(u,\xi) \, \sigma_c(c(u,\xi)) \rvert_1 + \\
	& +\lvert \alpha\sss{o}(u,\xi) \,\,\sigma_c \big(\alpha\sss{f}(u,\xi) x + \alpha\sss{i}(u,\xi)\, \sigma_c(c(u,\xi))\,\big) \,\rvert_1 \, - |x|\sss{1}-|\xi|\sss{1} .
	\end{aligned}
\end{equation*}
In view of the Lipschitzianity of the activation function $\sigma_c(\cdot)$, it holds that $\sigma_c(|v|\sss{1}) \leq |v|\sss{1}$, thus
\begin{align*}
V(\chi\uss{+}) \! - \!  V(\chi) \! \leq & \big[ 1+|\alpha\sss{o}(u,\xi)|\sss{1} \big] |\alpha\sss{f}(u,\xi)|\sss{1}|x|\sss{1} \! +\! \big[ 1+|\alpha\sss{o}(u,\xi)|\sss{1} \big] |\alpha\sss{i}(u,\xi)|\sss{1} |c(u,\xi)|\sss{1} \! - \!  |x|\sss{1} \! - \! |\xi|\sss{1} .
\end{align*}
Recalling the definition of $c(u,\xi)$ in \eqref{eq:aux_def}, it holds that
\begin{equation} \label{eq:delta_vchi}
\begin{aligned}
	V(\chi\uss{+}) \! - \! V(\chi) \!\leq & \big\{ [ 1 \! + \!|\alpha\sss{o}(u,\xi)|\sss{1} ] \, |\alpha\sss{f}(u,\xi)|\sss{1} \! - \! 1 \big\} |x|\sss{1} \! + \! \big\{ [ 1 \! + \! |\alpha\sss{o}(u,\xi)|\sss{1} ] \, |\alpha\sss{i}(u,\xi)|\sss{1}\, |U_c|\sss{1} \! - \! 1 \big\} |\xi|\sss{1}  \\
	& \,\, + \big\{ [1 \! + \! |\alpha\sss{o}(u,\xi)|\sss{1}] \, |\alpha\sss{i}(u,\xi)|\sss{1} |W_c|\sss{1} \big\} |u|\sss{1} + \big\{ [1 \! + \! |\alpha\sss{o}(u,\xi)|\sss{1}] \,|\alpha\sss{i}(u,\xi)|\sss{1} \big\} |b_c|\sss{1}.
\end{aligned}
\end{equation}
Let $\zeta=[u^T,\xi^T,\mathbf{1}_{n_x}^T]^T$, where $\mathbf{1}_{n_x}$ is the unitary column vector with length $n_x$. From \eqref{eq:bounds}, it follows that $\zeta \in[-1,1]^{2n_x+n_u}$.
By means of this, we can provide an upper bound for $|\alpha\sss{f}(u,\xi)|\sss{1}$:
\begin{equation*}
	|\alpha\sss{f}(u,\xi)|\sss{1} \! \leq \max_{\zeta} | {\rm diag}\left(\sigma_g\!\left( \left[ W_f \,\, U_f \,\, b_f \right] \zeta \right)\right) \! |\sss{1} \! = \!  \max_{\zeta} | \sigma_g \! \left( \left[ W_f \,\, U_f \,\, b_f \right] \zeta \right) \! |\sss{\infty} \! = \! \sigma_g \! \left( |  \left[ W_f \,\, U_f \,\, b_f \right]  \!|\sss{\infty} \! \right).
\end{equation*}
Similarly,
\begin{equation*}
\begin{aligned}
	|\alpha\sss{o}(u,\xi)|\sss{1} \leq & \,\sigma_g\left( |  \left[ W_o \,\, U_o \,\, b_o \right] |\sss{\infty} \right), \\
	|\alpha\sss{i}(u,\xi)|\sss{1} \leq & \, \sigma_g\left( |  \left[ W_i \,\,\, U_i \,\,\, b_i \right] |\sss{\infty} \right) .
\end{aligned}
\end{equation*}
In view of these bounds, conditions \eqref{cond_teo1} ensure the existence of a strictly positive scalar $\delta$ such that
\begin{equation} \label{eq:eps_bound}
	\begin{aligned}
	\left( 1+|\alpha\sss{o}(u,\xi)|\sss{1} \right) \, &|\alpha\sss{f}(u,\xi)|\sss{1} - 1 \leq -\delta, \\
	\left( 1+|\alpha\sss{o}(u,\xi)|\sss{1} \right) \, &|\alpha\sss{i}(u,\xi)|\sss{1} \,|U_c|\sss{1} -1 \leq -\delta,
	\end{aligned}
\end{equation}
for any $\xi\in\Xi$. Moreover, in light of \eqref{eq:bound_sigmag}, it follows that $|\alpha\sss{f}(u,\xi)|\sss{1}$, $|\alpha\sss{o}(u,\xi)|\sss{1}$ and $|\alpha\sss{i}(u,\xi)|\sss{1}$ are upper-bounded by $1$.
Combining \eqref{eq:delta_vchi} and \eqref{eq:eps_bound}, and since $ [1+|\alpha\sss{o}(u,\xi)|\sss{1}] | \, \alpha\sss{i}(u,\xi)|\sss{1} \, |W_c|\sss{1} \leq 2|W_c|\sss{1}$ and $[1+|\alpha\sss{o}(u,\xi)|\sss{1}] |\alpha\sss{i}(u,\xi)|\sss{1} \, |b_c|\sss{1}\leq 2|b_c|_1$, it is possible to derive that
\begin{equation*}
	V(\chi\uss{+}) - V(\chi) \leq - \delta |x|\sss{1} - \delta |\xi|\sss{1} + 2 |W_c|\sss{1} |u|\sss{1} + 2 |b_c|\sss{1} \leq - \delta |\chi|\sss{2} + 2|W_c|\sss{1}\sqrt{n_u}|u|\sss{2} + 2\sqrt{n_x}|b_c|\sss{2} .
\end{equation*}
In view of these results, it is possible to conclude that conditions \eqref{secondCond1} are satisfied with
\begin{align*}
	-\psi(|\chi|\sss{2})&= - \delta \, |\chi|\sss{2}, \\
	\sigma_u(|u|\sss{2})&= 2|W_c|\sss{1}\sqrt{n_u} \,|u|\sss{2}, \\
	\sigma_b(|b_c|\sss{2})&= 2\sqrt{n_x}\,|b_c|\sss{2}.
\end{align*}
System \eqref{eq:LPV} admits a time invariant ISS-Lyapunov function and, in view of Theorem \ref{Teo:ISSLyap}, it is ISS. 
\hfill $\blacksquare$

\clearpage
\bibliography{lstmbiblio.bib}

\end{document}